\documentstyle[preprint,aps]{revtex}
\begin{document}

\draft

\title{Non-Fermi liquid behavior in an extended Anderson model}

\author{Yu-Liang Liu,$^a$ Zhao-Bin Su,$^{a,b}$ and Lu Yu$^{a.b}$}
\address{$^a$International Center for Theoretical Physics, P. O. Box 586, 34100
 Trieste, Italy,\\
$^b$Institute of Theoretical Physics, CAS, 100080 Beijing, China.}

\maketitle

\begin{abstract}

An extended Anderson model, including screening channels 
(non-hybridizing, but interacting with the local orbit), is studied within 
the Anderson-Yuval approach, originally devised for the single-channel 
Kondo problem. By comparing the perturbation expansions of this model 
and a generalized resonant level model, the spin-spin correlation functions 
are calculated which show non-Fermi liquid exponent depending on the 
strength of the scattering potential. The relevance of this result to 
experiments in some heavy fermion systems is briefly discussed.

\end{abstract}
\vspace{1cm}

\pacs{75.20.Hr, 75.30.Mb, 71.28.+d, 71.27.+a}

\newpage

\section{Introduction}

The Anderson model, originally proposed to explain the formation of the 
local magnetic moment in metals, has played a fundamental role in 
exploring the correlation effects in many-body systems. Various 
theoretical approaches, including the Hartree-Fock\cite{pw1}, scaling 
theory\cite{pw2}, numerical renormalization group\cite{hrk}, perturbation 
expansion\cite{ky}, variational calculations\cite{cm1}, slave-boson mean 
field theory\cite{pc}, large-N diagrammatic expansion\cite{hk}, exact 
Bethe-Ansatz solution\cite{pwie1}, and many others\cite{for}, have been applied
to study this important issue. The general 
understanding has been that the Anderson model, in both electron-hole 
symmetric and asymmetric cases (including the valence fluctuation regime) 
exhibits a local Fermi liquid (FL) behavior, i.e., the impurity 
contribution to the specific heat and magnetic susceptibility is regular.

Recently, this "common understanding" has been questioned by studies of a 
 generalized Anderson model, including the so-called screening 
channels\cite{ip,cs,gm1,gm2}. In fact, it was realized long time ago by 
Anderson and Haldane that the original Anderson model is not complete to 
fulfil the local charge neutrality Friedel sum rule in the mixed-valence 
regime, $i.e.$, to satisfy  two different equations, 
corresponding to two different valence states by varying only one single 
parameter--the local state energy level\cite{fdm}. 
In addition to the channel of 
conduction electrons which hybridize with the local state, "screening" 
channels have been included, which do not hybridize with the local state 
but are related to it via Coulomb interactions. Haldane has also presented a 
Hartree-Fock mean field theory for this extended model\cite{fdm},
which, unfortunately, missed the non-FL behavior contained in it.

Recently, Varma and collaborators\cite{ip,cs} have revived interest to 
this model, considering it as a single site version of the three-band 
Hubbard-type model proposed to describe the cuprates\cite{cm2}. The 
"screening" channels correspond to oxygen orbits, not mixing with 
copper due to symmetry. Their original motivation was to find a 
microscopic justification for the phenomenological marginal FL 
theory\cite{cm3}. At first, some numerical evidence of non-FL behavior 
was provided by a Wilson renormalization group study\cite{ip}. Later, a 
strong coupling limit Hamiltonian was derived which exhibits a quantum 
critical point separating the Kondo and empty-orbit 
regions\cite{cs,gm1}. The correlation functions around this critical 
point show non-FL behavior. However, the calculation of the correlation 
functions is very delicate due to the presence of the single occupancy 
constraint. We believe the correct result has been obtained in \cite{gm2} 
which 
demonstrates a power law singularity for the impurity specific heat and 
magnetic susceptibility, namely, 
$C_{imp}/C_{0}\sim\chi_{imp}/\chi_{0}\sim T^{-3/4}$ in the unitarity 
limit (phase shift $=\pi/2$), 
in contrast to the logarithmic singularity anticipated 
earlier\cite{cs}. Whether this result is relevant for the high $T_{c}$ 
materials, is an open question. However, some heavy fermion compounds show 
such type of singularities. In particular, the alloys $UPd_{x}Cu_{5-x} 
\;(x=1, 1.5)$ have been studied in detail, using both static and neutron 
scattering techniques\cite{fora}. The critical exponent extracted from 
the experimental data\cite{mc1,mc2} is $\Delta=1/3$, which is qualitatively 
consistent 
but quantitatively different from the theoretical result obtained for the 
generalized Anderson model\cite{gm2}.

The calculations of \cite{gm2} have been carried out using the bosonization 
technique and the canonical transformations which provide values for 
the critical exponents only in the unitarity limit. 
In view of the importance of this issue and the need to obtain results 
away from the unitary limit \cite{ts} it is highly desirable 
to have another way to 
reconfirm and extend the previous result\cite{gm2}. In this paper we 
apply the Anderson-Yuval (AY) approach\cite{gy} to consider the extended 
Anderson model.

The AY approach\cite{gy} was originally devised to study the single 
channel Kondo problem. The main idea of this approach is based on the 
time-dependent one-body formulation of the X-ray edge problem, developed 
by Nozi\`{e}res and De Dominicis (ND) \cite{pn}. Using the AY perturbation 
expansion Toulouse could map the Kondo problem onto a resonant level 
model from which he could derive the well-known strong-coupling Toulouse 
limit\cite{gt}. Later, the bosonization technique was used by 
Schlottmann\cite{ps} to calculate physical properties in this strong 
coupling limit, although  bosonization was employed much earlier by 
Schotte to derive the AY perturbation expansion for the partition 
function \cite{kds}. The equivalence of this perturbation expansion with 
that of a resonant level model (at and away from the Toulouse limit) was 
shown explicitly and made use of for studying the physical properties by 
Wiegmann and Finkel'stein\cite{pwie2}. Recently, Fabrizio, Gogolin and 
Nozi\`{e}res \cite{mf} have generalized the AY approach to consider 
the asymmetric two-channel Kondo problem and the FL-non-FL crossover within
 that model. They have mapped term-by-term the perturbation expansion 
of the two-channel Kondo problem to that of a generalized resonant level 
model. Under the scaling assumption these authors could provide an 
analytical description of the FL-non-FL crossover. In this paper we will 
follow their approach rather closely. Bosonization and AY approach are 
similar in the sense of scaling assumption, but the latter is not limited 
to the Born approximation and can be used away from the unitarity limit.
Moreover, it is a "brute force" partial resummation of diagrams without
 resorting to canonical transformations used in the bosonization approach.

We should mention that there is another generalization of the Anderson 
model considered by Si and Kotliar\cite{qs,gk}, who have included all 
possible density-density interactions of the hybridizing channel without 
invoking the screening channels. Using the renormalization group 
expansion they considered the weak coupling case\cite{qs}, while the 
strong coupling limit was treated by bosonization\cite{gk}. In the final 
section we will discuss how their results are related to ours in the 
overlapping region.

The rest of the paper is organized as follows: The model  is
defined and the basic strategy to treat the model is given in Sec.II, while
the perturbation expansion and the mapping onto a generalized resonant-level
model is presented in Sec.III. Furthermore, the correlation functions
are calculated in Sec. IV to be confronted with experimental results.
Finally, some concluding remarks are made in Sec.V.

\section{The model and the basic strategy}

The Hamiltonian we consider in this paper is given as a sum
$$ H_{T}=H+\bar{H},$$
where the hybridizing part is
\begin{eqnarray}
&& 
H=H_{0}+H_{I}+t\sum_{\sigma}(c^{+}_{\sigma}(0)d_{\sigma}+
d^{+}_{\sigma}c_{\sigma}(0)),\nonumber\\&&\hspace{.2cm}
H_{0}=\sum_{\sigma,k}\epsilon_{k}c^{+}_{k\sigma}c_{k\sigma},
\nonumber\\&&\hspace{.2cm}
H_{I}=V_{0}\sum_{\sigma}c^{+}_{\sigma}(0)c_{\sigma}(0)(n_{\sigma}-\frac{1}{2}),
\label{A} 
\end{eqnarray}\narrowtext\noindent
while the screening part is
\begin{equation}
\bar{H}=\sum_{l,k,\sigma}\epsilon_{k}c^{+}_{lk\sigma}c_{lk\sigma}
+\sum_{l,\sigma}V_{l}c^{+}_{l\sigma}(0)c_{l\sigma}(0)(n-\frac{1}{2}).
\label{B}
\end{equation}
Here $c^{+}_{k\sigma},\; c_{k\sigma}$ are conduction electron operators 
in the hybridizing channel, with $c^{+}_{\sigma}(0),\; c_{\sigma}(0)$
as their Fourier transforms at the origin of the coordinates (where the 
impurity is located). Similarly, $c^{+}_{lk\sigma},\; c_{lk\sigma}$ are 
conduction electron operators in the screening channels. 
$n=\sum_{\sigma}n_{\sigma},\; n_{\sigma}=d^{+}_{\sigma}d_{\sigma}$ is the 
local impurity operator. The Hubbard $U$ on impurity itself has been 
taken as infinity, so the only allowed states are 
$|0>,\;|\sigma>,\;\sigma=\uparrow,\; \downarrow$. $V_{0},\;V_{l}$ are the 
Coulomb interactions of the local electron with conduction electrons in 
the hybridizing and screening channels, respectively. The essential part 
of this Hamiltonian is the same as that in earlier 
papers\cite{cs,gm1,gm2}. We have not included here the anti-parallel spin 
Coulomb interactions 
$V^{'}_{0}\sum_{\sigma}c^{+}
_{\sigma}c_{\sigma}(n_{\bar{\sigma}}-\frac{1}{2})$ 
and the spin-flip scattering 
$V_{\bot}\sum_{\sigma}c^{+}_{\sigma}c_{\bar{\sigma}}d^{+}_{\bar{\sigma}}
d_{\sigma}$ in the hybridizing channel, considered in \cite{gm1,gm2}.
As shown in those references, these terms do not affect essentially the 
behavior in the strong coupling limit. For simplicity, we have also taken 
the local level $\epsilon_{d}$ at the Fermi level, since we are mainly 
interested in the critical behavior itself, rather than the level 
renormalization per se.

Before proceeding, let us briefly recall the basic strategy of 
ND\cite{pn} and AY\cite{gy} to see why their approach can be applied to 
our problem. The crucial term in the X-ray edge problem is $V\sum 
c^{+}(0)c(0)dd^{+}$, where $d$ is the deep-level electron, while 
$c^{+}(0),\;c(0)$ are the conduction electron operators at the origin. As it
stands, this is a many-body problem. However, ND have realized that it can 
be converted into a time-dependent one-body problem, because the 
scattering potential is effective only after X-ray absorption or before 
X-ray emission (when $dd^{+}=1$). Since the internal degrees of freedom 
for the local electron are not involved, its propagator in the standard 
many-body technique can be traced out, thus converting it into a one-body 
problem. Moreover, instead of the usual Fourier representation of Green's 
functions, ND solved the Dyson equation directly in the time domain. The 
integral equation then obtained turned out to be singular, of the 
Muskhelishvili-type\cite{nim} from which the power-law time asymptotic 
behavior is extracted with exponents expressed in terms of  a phase shift 
$\delta=\tan^{-1}(\pi\nu_{0}V)$, where $\nu_{0}$ is the conduction electron 
density of states.

Soon afterwards AY\cite{gy} realized that the ND trick can be used to obtain 
the perturbation expansion for the Kondo problem with a Hamiltonian
\begin{equation}
H_{K}=\frac{1}{2}J\sum[c^{+}_{\uparrow}c_{\downarrow}S^{-}
+c^{+}_{\downarrow}c_{\uparrow}S^{+}+(c^{+}_{\uparrow}c_{\uparrow}-
c^{+}_{\downarrow}c_{\downarrow})S^{z}],
\label{C}
\end{equation}
where, $S^{+},\;S^{-},\;S^{z}$ are local spin components, while 
$c^{+}_{\uparrow},\;c_{\downarrow},\;...$ are conduction electron 
operators at the origin. For any given sequence of spin flips at moments
$t_{1},\;...,\;t_{n}$ (from up to down), $t_{1}^{'}, \; ...,\; t^{'}_{n}$ 
(from down to up) in a perturbation expansion of the partition 
function, the asymptotic expression can be used if the time difference 
$t_{i}-t^{'}_{i}, \; ...$, is much greater than the transient time $t_{0}\sim$ 
the inverse of the bandwidth. The exponents can again be expressed in 
terms of the phase shift 
$\delta=\delta_{+}-\delta_{-}=2\tan^{-1}\frac{\pi\nu J}{4}$, due to the 
difference of scattering potential experienced by the up and down spins. By 
mapping this perturbation onto an expansion for some kind of 
resonant level model which can be solved exactly in limiting cases,  the 
low energy physics can be extracted. This was the basic strategy of 
\cite{pwie2} and \cite{mf} and will be followed in this paper. Instead of 
$|\uparrow>$ and $|\downarrow>$ two states for the single channel Kondo
problem, we have here three states $|0>, \; |\uparrow>,\;$  and $\; 
|\downarrow>$. Also, we have screening channels in addition to the 
hybridizing channels. Nevertheless, the above programme can be still 
implemented, as seen from the next Section.

\section{Perturbation expansion and mapping onto a generalized resonant 
level model}

Like in the X-ray edge and Kondo problems, we are interested in the time 
evolution of the system described by
$$F(\tau)=<0|e^{iH_{T}\tau}|0>,$$
where $H_{T}$ is the total Hamiltonian given by (\ref{A}) and (\ref{B}), while 
$|0>$ is the ground state which is degenerate in the mixed valence regime 
with the local electron in one of the states $|\alpha>,$ $ \; \alpha=0, 
\;\uparrow \; and \; \downarrow $. We consider the perturbation expansion 
in terms of the hybridization parameter $t$
\begin{eqnarray}
&& 
F(\tau)=<0|e^{iH^{'}\tau}T\{ e^{i
\int^{\tau}_{0}d\tau^{'}H_{h}(\tau^{'})}\}|0>
=\sum^{\infty}_{n=0}\int^{\tau}_{0}d\tau_{2n}\int^{\tau_{2n}}_{0}d
\tau_{2n-1}\cdots\int^{\tau_{2}}_{0}d\tau_{1}\nonumber\\
&&\hspace{1cm}
<0|e^{iH^{'}(\tau-\tau_{2n})}iH_{h}\cdots e^{iH^{'}(\tau_{2j+1}-\tau_{2j})}
iH_{h} e^{iH^{'}(\tau_{2j}-\tau_{2j-1})}iH_{h}\cdots
iH_{h}e^{iH{'}\tau_{1}}|0>
\label{D}\end{eqnarray}
\narrowtext\noindent with
\begin{eqnarray}
&& H_{h}=t\sum_{\sigma}(c^{+}_{\sigma}(0)d_{\sigma}+d^{+}_{\sigma}c_{\sigma}
(0)),\nonumber\\&&\hspace{.2cm}
H^{'}=H_{0}+H_{I}+\bar{H}.
\label{E}\end{eqnarray}
\narrowtext\noindent
Due to the presence of $d^{+}_{\sigma},\; d_{\sigma}$ in $H_{h}$, only 
even order terms are kept.
A typical term will contain either  spin up (down) operators only, or 
mixed. However, in view of the single occupancy constraint the spin up 
and down states can be connected only via the empty state. Therefore, 
these terms can be separated into up and down blocks. Using the relations
\begin{equation}
 d^{+}_{\sigma}e^{iH^{'}\tau_{j}}|\alpha>=\left\{\begin{array}{ll}
e^{iH^{'}_{\alpha}\tau_{j}}|\sigma>, &\mbox{if $|\alpha>=|0>,$}\\
0, &\mbox{if $|\alpha>\neq |0>,$}\end{array}\right.\nonumber
\end{equation}
\begin{equation}
d_{\sigma}e^{iH^{'}\tau_{j}}|\alpha>=\left\{
\begin{array}{ll}
0, &\mbox{if $|\alpha>\neq |\sigma>$},\\
e^{iH^{'}_{\alpha}\tau_{j}}|0>, &\mbox{if $|\alpha>=|\sigma>$}
\end{array}\right.
\label{F}
\end{equation}\narrowtext\noindent
with $H^{'}_{\alpha}\equiv<\alpha|H^{'}|\alpha>$,  we can trace out 
$d_{\sigma}$ operator. Since the Hamiltonian $H^{'}_{\alpha}$ is 
different for the  neighboring time moments $\tau_{j}$ and $\tau_{j+1}$ (or 
$\tau_{j-1}$), the calculation is similar to the X-ray edge or Kondo 
problem and we can apply AY approach to handle it. Explicitly, we fix 
$\tau_{1j}\;(\tau^{'}_{1j})$ as moments
when a local spin-up electron is created (annihilated). Likewise 
$\tau_{2j}\;(\tau^{'}_{2j})$ for local spin-down electron. As said 
before, due to the single occupancy constraint, the only allowed 
sequences are: 
$(\tau_{1j},\;\tau_{1j}^{'}),\;(\tau_{2j},\;\tau_{2j}^{'})$,
i.e., the up and down sequences do not intersect each other. 

In the absence of the scattering potentials $V_{0}=V_{l}=0$, the 
propagator for the conduction electron in the local presentation
$G_{\sigma}(\tau)=<Tc_{\sigma}(\tau)c^{+}_{\sigma}(0)>$ is\cite{pn}:
\begin{equation}
G^{(0)}(\tau)=\frac{i\nu_{0}}{\tau-i\xi^{-1}_{0}{\rm sgn}\tau},
\label{G}
\end{equation}
where $\xi_{0}$ is the bandwidth, serving as a cut-off. The contribution 
of $2n$ vertices connecting $\tau_{\sigma i},\;\tau^{'}_{\sigma j}\;
(\sigma=1,2)$ will be\cite{gy,mf}:
\begin{equation}
D_{\sigma}=\frac{\displaystyle\prod_{i<j}(\tau_{\sigma i}-\tau_{\sigma j})
\displaystyle\prod_{i<j}(\tau^{'}_{\sigma i}-\tau^{'}_{\sigma j})}{
\displaystyle\prod_{ij}(\tau_{\sigma i}-\tau^{'}_{\sigma j})}.
\label{I}
\end{equation}

In the absence of the scattering $V_{0}=V_{l}=0$, the contribution of 
separate terms in (\ref{D}) will be
\begin{equation}
U_{0}=D_{\uparrow}D_{\downarrow}.
\label{J}
\end{equation}

Now we turn on the scattering potential. First consider the hybridizing 
channel. For the spin up conduction electron at moment $\tau_{1i}$ a 
scattering potential is switched on to give a phase shift
\begin{equation}
\delta=2\delta_{0}=2\tan^{-1}(\frac{\pi\nu_{0}V_{0}}{2}),
\label{K}
\end{equation}
while at moment $\tau^{'}_{1i}$ an opposite phase shift is produced. On 
the other hand, the spin down conduction 
electron does not experience any phase 
shift at these moments. Of course, at moments $\tau_{2i}$ and
 $\tau_{2i}^{'}$  the situation is reversed.
There are two types of contributions to the renormalization of the 
conduction electron propagation: from an open line $U_{L}$ and closed 
loops (vacuum fluctuations). As shown earlier\cite{gy,mf},
\begin{eqnarray}
&&U_{L}=(D_{\uparrow}D_{\downarrow})^{-2\delta/\pi},
\nonumber\\&&\hspace{.2cm}
U_{c}=(D_{\uparrow}D_{\downarrow})^{\delta^{2}/\pi^{2}}.
\label{L}
\end{eqnarray}\narrowtext\noindent
In deriving (\ref{L}) we have taken into account the fact that in the 
process $|0>\rightarrow|\uparrow>$, nothing is changed for 
$H^{'}_{\alpha},$ with $\alpha=0,\;\downarrow$, so there are no crossing terms.

Next we consider the screening channels. For channel $l$ the conduction 
electron gets a phase shift
\begin{equation}
\delta_{l}=2\delta(0)_{l}=2\tan^{-1}(\frac{\pi\nu_{0}V_{l}}{2})
\label{M}
\end{equation}
at both $\tau_{1i}$ and $\tau_{2i}$, and an opposite phase shift at 
$\tau^{'}_{1i}$ and $\tau^{'}_{2i}$. The screening electrons contribute 
only to the closed loops. However, as seen from (\ref{B}), the screening 
channel electrons scatter on the total charge 
$n=\sum_{\sigma}n_{\sigma}$, so the process $|\uparrow>\rightarrow|0>$ 
will affect also spin down conduction electron in the screening channels. 
Therefore we will have crossing terms which are similar to the case of 
multichannel Kondo model\cite{mf} with the spin index replacing the 
channel index there. The final result is
\begin{equation}
U_{cl}=(D_{\uparrow}D_{\downarrow}F)^{2(\frac{\delta_{l}}{\pi})^{2}}
\label{N}
\end{equation}
with the crossing term
\begin{equation}
F=\frac{\displaystyle\prod_{ij}(\tau_{1i}-\tau_{2j})(\tau^{'}_{1i}
-\tau^{'}_{2j})}
{\displaystyle\prod_{ij}(\tau_{1i}-\tau^{'}_{2j})(\tau^{'}_{1i}-\tau_{2j})}.
\label{P}
\end{equation}

Putting everything together, we find the total contribution of all 
channels to a given term is
\begin{equation}
U_{T}=U_{0}U_{L}U_{c}\prod^{N}_{l=1}U_{cl}
=(D_{\uparrow}D_{\downarrow})^{(1-\frac{\delta}{\pi})^{2}}
(D_{\uparrow}D_{\downarrow}F)^{2\sum^{N}_{l=1}(\frac{\delta_{l}}{\pi}
)^{2}}.\label{Q}
\end{equation}\narrowtext\noindent

Following earlier treatments\cite{pwie2,mf}, we now consider a 
generalized resonant level model
\begin{equation}
H=H_{0}+\lambda[(d^{+}_{\uparrow}+d^{+}_{\downarrow})\psi(0)+
\psi^{+}(0)(d_{\uparrow}+d_{\downarrow})] + V\psi^{+}(0)\psi(0)
(n-\frac{1}{2}),
\label{S}
\end{equation}
where 
$n=(d^{+}_{\uparrow}+d^{+}_{\downarrow})(d_{\uparrow}+d_{\downarrow})$, 
with constraint $d^+_{\uparrow}d_{\uparrow} + d^+_{\downarrow}d_{\downarrow}
\le 1$, $\psi^{+}(0),\;\psi(0)$ a spinless fermion. 
As before, we can expand the evolution operator in terms of $\lambda$. 
Again only even order terms survive and there are three types of terms 
(containing only 
$d^{+}_{\uparrow},
\;d_{\uparrow}\;or\;d^{+}_{\downarrow},\;d_{\downarrow}$, or mixed. 
Also, due to the single occupancy constraint any $ |\sigma>$ state can 
be created only from the empty state. We can then repeat the same procedure to 
derive contributions from the open line and closed loops. However, there 
is one important difference, namely, the single type of spinless fermion 
$\psi(0)$ is coupled to both $d_{\uparrow}\;$ and $\;d_{\downarrow}$, so there 
are crossing terms even for the free propagator (where $V=0$). When we 
switch on the scattering potential at moments $\tau_{\sigma i}$ (the 
process $|0>\rightarrow |\sigma>$) $\psi$ gets a phase shift
\begin{equation}
\delta^{'}=2\tan^{-1}(\frac{\pi\nu_{0}V}{2})
\label{T}
\end{equation}
and an opposite one at moments $\tau^{'}_{\sigma i}$ during the process 
$|\sigma>\rightarrow |0>$. Summing up the contributions from the free 
propagator, the open line and the closed loops, we find the $n$-th order 
term of $\lambda$ is given by
\begin{equation}
U^{'}=(D_{\uparrow}D_{\downarrow}F)^{(1-\frac{\delta^{'}}{\pi})^{2}}.
\label{W}
\end{equation}

If we take
\begin{equation}
\delta=\pi,\;\;\;\; 2\sum^{N}_{l=1}(\frac{\delta_{l}}{\pi})^{2}
=(1-\frac{\delta^{'}}{\pi})^{2}
\label{V}
\end{equation}
the extended Anderson model $H_{T}$, given by (\ref{A}) and (\ref{B}), and 
the generalized resonant-level model (\ref{S}) are equivalent to each other 
via a term-by-term mapping of the perturbation expansion. It is expected 
that they should contain the same low-energy physics. Of course, there is 
an underlying assumption that $V$ is the only scaling parameter for this 
universality class. A similar assumption was made for the single channel 
Kondo and the two-channel Kondo problem with channel asymmetry\cite{mf}. 

\section{Correlation functions and physical implications}

Before proceeding we compare first the result of the preceding Section
with earlier calculations using bosonization\cite{gm1,gm2}. In the 
previous work a model very similar to (\ref{A}), (\ref{B}) was considered with 
additional opposite spin Couloumb interaction 
$V^{'}_{0}\sum_{\sigma}c^{+}
_{\sigma}c_{\sigma}(n_{\bar{\sigma}}-\frac{1}{2})$ 
and spin-flip scattering 
$V_{\bot}\sum_{\sigma}c^{+}_{\sigma}c_{\bar{\sigma}}d^{+}_{\bar{\sigma}}
d_{\sigma}$ in the hybridizing channel (see Eq. (6) in \cite{gm2}). It 
was shown there by using the canonical transformation that the strong 
coupling Toulouse limit is reached for $V_{0}\rightarrow\infty, 
\;V_{0}^{'}\rightarrow 0$, 
$\tilde{V}_{s}=\sqrt{2N_{s}}V_{s}\rightarrow\infty$, where $V_{s}$ is 
the Couloumb potential in the screening channel and $N_{s}$ is the number 
of channels. As follows from (\ref{K}), (\ref{M}), (\ref{T}) and (\ref{V}), we 
have  the same Toulouse limit here. In fact, 
$\delta_{0}=\tan^{-1}(\pi\nu_{0}V/2)=\frac{\pi}{2},\; \delta^{'}=0, \;
2\sum^{N}_{l=1}(\frac{\delta_{l}}{\pi})^{2}=1$, meaning that the 
unitarity limit is reached in both hybridizing and screening channels. The 
earlier calculations correspond to the Born approximation of our present 
result summed to infinite orders. This reconfirms the consistency of 
bosonization and ND approach. However, here we have obtained the mapping to 
the generalized resonant-level model in a broader regime, namely, the 
unitarity limit should be reached in the hybridizing channel 
($\delta=\pi$), but not necessarily in the screening channels 
($\delta^{'}\neq 0$). Since the Toulouse limit is materialized at 
$V_{0}^{'}=0$, the opposite spin scattering is not essential. As seen in 
\cite{gm2}, the effect of the $V_{\bot}$ term is reflected only in the energy 
difference of $\alpha$ and $\beta$ particles 
($\alpha=\frac{1}{\sqrt{2}}(d_{\uparrow}+d_{\downarrow}),\;
\beta=\frac{1}{\sqrt{2}}(d_{\uparrow}-d_{\downarrow}))$ which we will 
include in our following discussion.

Now we study the physical properties of the generalized resonant-level 
model (\ref{S}) rewriten as
\begin{eqnarray}
&&H=H_{0}+H_{h}+H_{I},\nonumber\\&&\hspace{.2cm}
H_{0}=\sum_{k}\epsilon_{k}\psi^{+}_{k}\psi_{k}+\epsilon_{\alpha}
\alpha^{+}\alpha+\epsilon_{\beta}\beta^{+}\beta,
\nonumber\\&&\hspace{.2cm}
H_{h}=\lambda(\alpha^{+}\psi(0)+\psi^{+}(0)\alpha),
\nonumber\\&&\hspace{.2cm}
H_{I}=V\psi^{+}(0)\psi(0)(\alpha^{+}\alpha-\frac{1}{2})
\label{X}
\end{eqnarray}\narrowtext\noindent
which should still satisfy the single occupancy constraint
$\alpha^{+}\alpha+\beta^{+}\beta\leq 1$. First we calculate the 
scattering amplitude
\begin{eqnarray}&&
S(t)=<|e^{iHt}|>=exp\{C(t)\},
\nonumber\\&&\hspace{.2cm}
C(t)=<T\{exp[i\int^{t}_{0}d\tau(H_{h}(\tau)+H_{I}(\tau))]\}>_{c}.
\label{Y}
\end{eqnarray}\narrowtext\noindent
There are contributions from the spinless fermion $\psi$ and the $\alpha$ 
particle, both from the closed loops. Neglecting the energy level 
renormalization factors, the long time asymptotics are given by
\begin{eqnarray}
&& C(t) \sim -(\frac{\delta_{1}}{\pi})^{2}\ln t - (\frac{\delta_{2}}{\pi})^{2}
\ln t,\nonumber\\&&\hspace{.2cm}
S(t)\;\sim\;t^{-(\frac{\delta_{1}}{\pi})^{2}-(\frac{\delta_{2}}{\pi}
)^{2}}(1+e^{-i\epsilon_{\alpha}t}),
\label{Z}
\end{eqnarray}\narrowtext\noindent
where
\begin{equation}
\delta_{1}=\tan^{-1}(\pi\rho V),\;\;\;
\delta_{2}=\tan^{-1}(\frac{\pi\rho\lambda^{2}}{\epsilon_{\alpha}}).
\label{Z1}
\end{equation}
with $\rho$ as the density of states for the spinless fermion $\psi$.
Here we have taken the ground state energy $E_{0}=0$ and 
$\delta_{2}=\frac{\pi}{2}$, if $\epsilon_{\alpha}=0$.
There are crossing terms coming from $H_h$ and $H_I$ of Eq. (\ref{X}),
but they do not contribute to power law singularities in the
correlation functions.

Using the asymptotic form for $S(t)$, we can calculate the propagator
\begin{equation}
<\beta(t)\beta^{+}(0)>\sim e^{-i\epsilon_{\beta}t}
(t)^{-(\frac{\delta_{1}}{\pi})^{2}-(\frac{\delta_{2}}{\pi})^{2}}
\label{Z2}
\end{equation}\narrowtext\noindent
and the spin-spin correlation function
\begin{equation}
M(t)\sim <S^{z}(t)S^{z}(0)>\sim \cos[(\epsilon_{\alpha}-\epsilon_{\beta})t]
(t)^{-(\frac{\delta_{1}}{\pi})^{2}-(\frac{\delta_{2}}{\pi})^{2}},
\label{Z3}
\end{equation}
where $S^{z}=\frac{1}{2}(d^{+}_{\uparrow}d_{\uparrow}-d^{+}_{\downarrow}
d_{\downarrow})=\frac{1}{2}(\alpha^{+}\beta+\beta^{+}\alpha)$.
Assuming $\delta_{1}=0, \;\delta_{2}=\frac{\pi}{2}$, we recover the 
previous result, i.e., $M(t)\sim t^{-\frac{1}{4}}$. This is a nice and 
independent check of the correctness of calculations in \cite{gm2}.
 Moreover, here we 
have also obtained result away from the unitarity limit $\delta_{1}=0$.
Of course, the asymptotic form is valid only for small $\delta_{1}$, 
because the model (\ref{X}) cannot be solved exactly.

As mentioned in the Introduction, the neutron scattering data as well as 
the static measurements in $UPd_{x}Cu_{5-x}\;(x=1,1.5)$ show a  power 
law behavior of the spin-spin correlation function and the impurity 
contribution to the specific heat and magnetic 
susceptibility\cite{fora,mc1,mc2}. Using the conformal invariance for the 
impurity problem it has been argued that the available data are 
consistent with a critical exponent\cite{mc2} $\Delta=\frac{1}{3}$, while 
the value following from the bosonization calculation\cite{gm2} was 
$\Delta_{B}=\frac{3}{4}$ which is rather big compared with the experimental
value. The deviation from the 
unitarity limit $\delta_{1}=0$ will reduce this value. To fit the data 
we need to assume
$\delta_{1}=\pi\sqrt{5/12}$
which is still rather big to justify the applicability of our
asymptotic expansion. Nevertheless, the correction is in the right 
direction.

\section{Concluding remarks}

Using the ND and AY perturbation expansion and mapping onto a generalized 
resonant-level model we have reconfirmed and extended earlier results on 
non-FL behavior in an extended Anderson model with additional screening 
channels. The extension to regions away from the unitarity limit of 
screening electron scattering improves the agreement with experiment. 
However, in view of its physical implications this issue  should be 
further studied using other techniques.

As mentioned in the Introduction, the Anderson model has been also 
generalized in a different way\cite{qs,gk}. Without going into detailed 
comparison we briefly comment on the study of the Toulouse limit in that 
model\cite{gk}. Those authors have correctly pointed out the change of 
sign for the $V_{\bot}$ term upon bosonization. However, the correlation 
functions were calculated there using the mean field approximation in 
handling the constraint which missed the non-FL exponent. Of course, the 
physical consequences depend strongly on the positions of the 
renormalized levels $\epsilon_{\alpha}$ and $\epsilon_{\beta}$ (see eq. 
(\ref{Z3})). If the difference is big, the fast oscillation will suppress 
the power law component in the frequency response. Since the theoretical 
calculation of level renormalization is very difficult, we may count on 
experimental indication which seems to show the existence of remaining 
degeneracy. The situation here is similar to the two channel Kondo 
case\cite{mf}. There the decoupling of $d^{+}+d$ from the conduction 
electron was a signature of non-FL behavior which becomes more apparent 
in the Majorana fermion formulation\cite{pco}. Here the $\beta$ particle 
decouples, giving rise to a X-ray edge like singularity and a residual entropy,
as in the case of two-channel Kondo. Of course, this analogy is more 
mathematical than physical. The difference of non-FL behavior in the 
miulti-channel Kondo model and the extended Anderson model in the mixed valence 
regime, and their possible connections have to be further explored. 
A detailed comparison  of the scaling theory with numerical Wilson RG studies,
as well as a stability analysis of the strong coupling fixed point
in the extended Anderson remain outstanding issues.

	Finally, we would like to thank G.M. Zhang for an earlier collaboration 
on this project and M. Fabrizio for helpful discussions. Mobility within
Europe involved in this research project was partly sponsored by EEC,
through contract ERB CHR XCT 940438.

\end{document}